# New oil modified acrylic polymer for pH sensitive drug release: Experimental results and statistical analysis

N. PANJA, A. K. CHATTOPADHYAY

Fuel Cell Division, College of Engineering and Physical Sciences, University of Birmingham, Edgbaston B15 2TT, UK

and

Aston University, Nonlinearity and Complexity Research Group,
School of Engineering and Applied Science, Birmingham B4 7ET, UK
Email: a.k.chattopadhyay@aston.ac.uk

**Address for correspondence**: a.k.chattopadhyay@aston.ac.uk

**Abstract:** We report results of an experimental study, complemented by detailed statistical analysis of the experimental data, on the development of a more effective control method of drug delivery using a pH sensitive acrylic polymer. New copolymers based on acrylic acid and fatty acid are constructed from dodecyl castor oil and a tercopolymer based on methyl methacrylate, acrylic acid and acryl amide were prepared using this new approach. Water swelling characteristics of fatty acid, acrylic acid copolymer and tercopolymer respectively in acid and alkali solutions have been studied by a step-change method. The antibiotic drug cephalosporin and paracetamol have also been incorporated into the polymer blend through dissolution with the release of the antibiotic drug being evaluated in bacterial stain media and buffer solution. Our results show that the rate of release of paracetamol gets affected by the pH factor and also by the nature of polymer blend. Our experimental data have later been statistically analyzed to quantify the precise nature of polymer decay rates on the pH density of the relevant polymer solvents. The time evolution of the polymer decay rates indicate a marked transition from a linear to a strictly non-linear regime depending on the whether the chosen sample is a general copolymer (linear) or a tercopolymer (non-linear). Non-linear data extrapolation techniques have been used to make probabilistic predictions about the variation in weight percentages of retained polymers at all future times, thereby quantifying the degree of efficacy of the new method of drug delivery.

# Graphical abstract

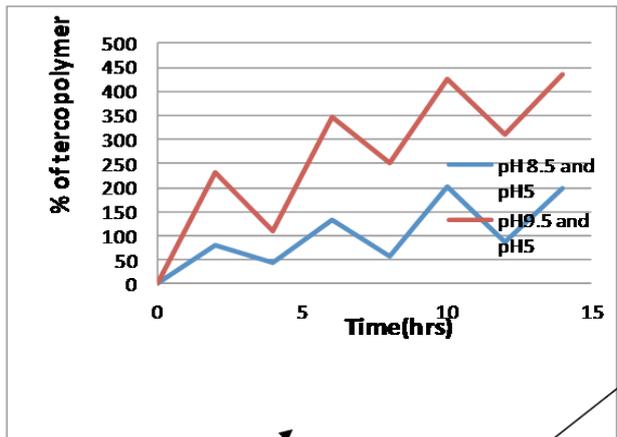
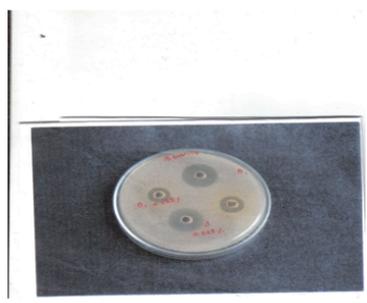
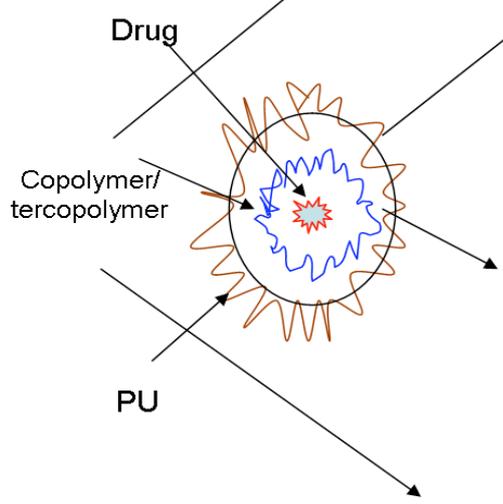
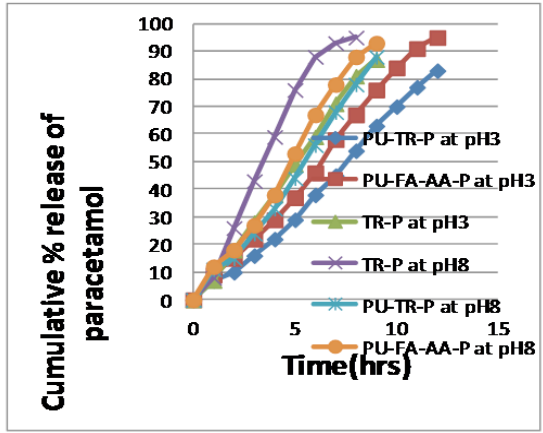
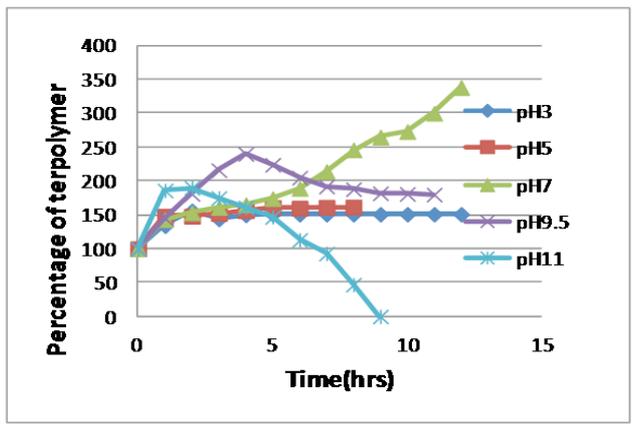

## 1. INTRODUCTION:

Synthesis and characterization of polymers for controlled drug release systems have assumed great importance among the polymer scientists in recent years due to their potential application in pharmaceutical industries. The biodegradable and water-soluble polymers have recently been used in controlled release drug formulation. The active chemical agents, such as drugs, are bound to the polymers via chemical and physical bonds. Thus the drugs are mixed in polymer solutions and the solid micro-sphere is prepared. These systems usually provide diffusion control release of drugs from the network structure or dissolve slowly in our body fluid. The release of drugs is controlled within a specified window that has a maximum value, defined by a limit of toxic level and a minimum value, below which the drug has minimal effect. The nature of polymers and there structure play important roles in controlling the release of drug from the drug bound polymer to body fluid and to the site of disease. Efforts are ongoing to prepare and modify industrially realizable synthetic polymers to achieve higher performance within a reasonably low cost portfolio.

Polymers for controlled drug delivery formulations are synthesized and studied for application and for better performance. The general preference is for polymers that are easy to process and are simultaneously reasonably inert in chemical solvents that are used in drug delivery formulations. Typical examples of such polymers are polymethymethacrylate, poly acrylic acid, polyacrylamaide, polyethylene or polyvinylacetate, poly methacrylic acid and polyurethane [1,2,3]. In recent years, polylactides ( PLA), polyglycolides (PGA) and polylactide-co-glycolides (PGLA) a have been extensively used in medical application and drug targeting [2,4,5 ]. These polymers have the advantages of biocompatibility and biodegradability that may be broken down into biologically compatible molecules while releasing active drug molecules that may be accelerated towards the site of action. The unwanted broken down molecules are removed from the human body via a metabolite route. Quality aspects of these polymers related to longevity and speed of delivery necessitate more testing and detailed study with respect to improvement in mechanical, oxidation degradation and processing properties [1,2].

H.Hchikawa et al [6] synthesized acrylic polymers that are thermo sensitive whose properties have been altered in response to ambient temperature. They surveyed the use of thermo sensitive acrylic polymers for control drug delivery system. Amongst a variety of synthesized acrylic polymers, some have been known to precipitate in water above a certain temperature while being water soluble below that temperature. When these polymers are physically or chemically cross-linked, they behave as hydrogels with reversible swelling or shrinking in response to temperature changes. Akin to our body temperature, the pH of the body fluid also changes and exhibits different thermal expansion coefficients in stomach and in intestine. In the recent years, biodegradable or bio-erodible polymers (hydrogels) have been increasingly used as candidates for pH sensitive polymers in controlled drug delivery devices because of their good and thermodynamically stable biocompatibility. Biodegradable hydrogels with three-dimensional network structure swell, but they do not dissolve in contact with water. Hydrogels are known to exhibit good biocompatibility and are highly responsive to external media such as temperature, pH, electrical pulses, etc. (Candy and Sharma [ 7], Chuang et al [ 8], Kim et al [ 9] Yazdani and Retuert [10].

Very recently (2005), Krishna Rao, et al [11] have developed a pH sensitive interpenetrating polymer network base on chitosan, acrylamide grafted-poly (vinyl alcohol) and hydrolyzed acrylamide – grafted polyvinyl alcohol that are crossed linked and are used in the control release of cefadroxil, an antibiotic drug. Blend of chitosan with other polymer have been reported in the earlier literature (Kurkuri and Aminabha [12,13,14], Miya, et al [15] and Nakatsuka and Andrady, et al [16] to study their reversible volume change properties in response to external stimuli such as pH or temperature. However, the main disadvantage of hydogels and cytosan are their poor mechanical properties caused by extensive swelling. Cakmakh, et al [17, 18] synthesized a graft polymer based on soya bean oil and methyl methacrylate and studied the biocompatibility of this sample in presence of bacterial adhesion. Polyurethane polymers are proven biocompatible and durable materials as recently shown by Bernacca, et al [19]. Amnon Sintov, et al [20] formulated lidocaine-polyurethane drug matrix to study the *in vitro* drug release and showed that the release of drugs was very slow. A remarkable property of double network hydrogels has recently been reported by Naficy, et al [21]; the tensile strength of such an agent experiences a dramatic change to 1/4-th its initial value when immersed in a polyethylene glycol solution with pH>4. A new direction in efficient drug delivery protocols in cancerous tumors has been advised by Du, et al [22] based on a nano-particle residue delivery system. An immunological version of the hydrogel method to study protein vaccines was employed by Crownover, et al [23] through radical polymerization of an endosomal releasing segment based on propylacrylic acid and a hydrophilic segment, although the method suffered from immune deficiencies in CD4+ and CD8+.

A recurring feature in many of the modern drug delivery mechanisms has been temperature modulations of pH factors and their follow-up effects in controlling delivery speeds [24-26]. Although quite evident in experimental reports [24-26], apart from rare exceptions [27], theoretical findings have remained few and far between. The present article, although not yet at the level of explicit modeling, is nevertheless an attempt to bridge this gap in between experimental and theoretical analyzes using a novel drug delivery method that affirms the time effectiveness of this delivery through detailed statistical analysis (detailed later).

Based on our literature survey, we found negligible evidence of oil modified polymers in controlled drug release, particularly on the development of unsaturated fatty acids-acrylic acid copolymer and tercopolymer (acryamide-acrylic acid-methyl methacrylate) and blend of polymers with polyurethane polymer matrices which have swelling and bio-erodible properties suitable for control drug release. In view of this, the present effort has been focused to prepare pH sensitive polymeric systems for drug delivery with better physiological compatibility, controlled release and mechanical properties with an advantage of low cost route to manufacture. In this work, we prepared two polymers. One copolymer was prepared from the unsaturated fatty acid and acrylic acid and another tercoplymer was prepared from acrylic acid, acryl amide and methyl methacrylate. Swelling and water soluble characteristics of copolymers were studied at different pH levels of water. The release of an antibiotic drug cephalosporin from the blend of PU resin and pH sensitive copolymer and tercopolymers was tested in culture media by the extent of inhibition of growth of the bacterial front. The release of paracetamol was tested by spectrophotometer.

## 2. MATERIALS AND METHODS:

All experimental protocols and methods adapted in this research comply with the particular recommendation of the parent institutions and all necessary approval of such protocols was obtained.

Synthesis of fatty acid and acrylic acid copolymer

The synthesis process involves three steps:

I. Preparation of fatty acid from dehydrated castor oil (DCO)

II. Preparation of fatty acid and acrylic acid copolymer

III. Preparation of tercopolymer

### 2.1 Preparation of fatty acid from dehydrated castor oil (DCO)

100 gm of dehydrated castor oil was taken in a 1 liter round bottomed flask in which 100 ml of 30% alcoholic NaOH was added and refluxed for 4 hours on water bath using water condenser. Air condenser was removed to evaporate methanol and then the entire solution was allowed to cool at room temperature for approximately 30 minutes.

Soap formed was in solid form; glycerol was separated from sodium salt of fatty acid by dissolving in hot water. This solution was taken in a separating funnel and acidified with 1:1 sulphuric acid ($H_2SO_4$) in the presence of methyl red as an indicator. Sulphuric acid was added till light pink color was developed in the glycerin layer at the lower part of the separating funnel. Upper part of the layer was fatty acid; it was separated and used for polymer synthesis.

### 2.2 Preparation of fatty acid and acrylic acid (FA-AA) copolymer

25 ml of acrylic acid was taken in a round-bottomed flask in which 25 ml of fatty acid, prepared from the above mentioned DCO process, was added. A solution mixture containing 0.25 gm potassium per sulfate (initiator), 0.25gm Na-hydrogen phosphate buffer, 0.25 gm Na-lauryl sulfate and 50 ml water was prepared and added to the mixture of fatty acid and acrylic acid in a round-bottomed flask. The mixture was heated in a water bath at about 55 $^0$C for 3 hours with slow stirring. Beyond a critical threshold value of the polymer viscosity, cooling stopped polymerization. The polymer was then tested for solubility and water absorption properties.

### 2.3 Preparation of acrylic acid, acryl amide, methyl mehacrylate (AA-AM-MMA) tercopolymer

20 ml acrylic acid and 10 gm of acryl amide were taken into an iodine flask and dissolved completely. A further 20 ml of methyl methacrylate (MMA) purified from hydroquinone was added to the above two monomers. The polymerization was carried out in the iodine tube, in presence of 0.25 gm of benzoyl peroxide catalyst, on the water bath at 65$^0$C for 2 hrs using a reflux condenser and a stirrer. Viscosity started to increase gradually and when the viscosity became very high, the temperature of the bath was removed from the assembly. The viscous polymer solidified on cooling.

### 2.4 Synthesis of polyurethane resin

1.1 mole of hexamethylene diisocyanate was combined with 1 mole proportion of polyethylene oxide glycol in a reaction mixer. Polymerization was carried out at 55 $^0$C with initial stirring. When polyurethane viscosity was raised to that suitable for a semi solid, 50 % of the polyurethane was taken into another beaker and mixed with 15% by weight of FA-AA copolymer and then the polyurethane blend was allowed to polymerize for a further 48 hours at isothermal condition.

2.5. **Drug loading**

i) 4 grams of tercopolymer was dissolved in dimethyl formamide and mixed with 25% by weight of paracetamol in a 250ml beaker. The drug mixture was then slowly evaporated.

ii) In another 250 ml beaker, polyurethane was mixed with 15% by weight tercopolymer respectively, followed by 25% by weight of paracetamol. This was then slowly evaporated.

iii) In the third 250ml beaker, polyurethane-FA-AA blend was dissolved in dimethyl formamide by stirring and was mixed with 25% by weight of paracetamol and then evaporated.

The drug loading of different polymeric blends with their formulation code is presented in Table 1.

Table 1: Drug loading and formula code

| Formulation | Percentage ratio | Code |
|---|---|---|
| PU-copolymer :Paracetamol | 75 :25 | PU-FA-AA-P |
| PU Tercopolymer : Paracetamol | 75 :25 | PU-TR-P |
| FA-AA :paracetamol | 75 :25 | FA- AA-P |
| Tercopolymer :Paracetamol | 75 :25 | TR-P |

**2.6 Solubility of copolymer (FA-AA) and tercoplymer (AA-AM-MMA) in organic solvents**

We performed the solubility test for both the fatty acid-acrylic acid (FA-AA) copolymer and tercopolymer (AA-AM-MMA) with different organic solvents. We found that the polymers were soluble in certain solvents while being insoluble in others (Table 2). Positive (+) signs indicate that the polymer is soluble and negative (-) signs indicate that the polymer is insoluble.

**Table 2: Solubility of copolymer (FA-AA) and tercoplymer (AA-AM-MMA) in organic solvents**

| Solvents | Polymers | |
|---|---|---|
| | F.A-A.A copolymer | Tercopolymer |
| Water | + | - |
| Acetone | + | + |
| Methanol | + | - |
| Xylene | - | - |
| 10% NaOH | + | + |
| Iso propanol | - | - |
| Toluene | - | - |
| Benzene | - | - |
| Carbon tetrachloride | - | - |
| Diethyl ether | - | - |
| Tetra hydro furan | + | + |
| Butyl acetate | - | - |
| Ethanol | + | + |
| Chloroform | - | - |
| Acetic acid | - | - |

**2.7 Swelling solubility characteristics of Copolymer (FA-AA)**

Water absorption and solubility characteristics of copolymer (FA-AA) were studied in tap water (pH 7.3), hydrochloric acid (HCl) solution (pH 5) and sodium hydroxide (pH 9.5). Fatty acid-acrylic acid copolymers weighing 2 grams were taken in a beaker and 25 ml of tap water (pH 7.3) was added and kept for ½ hr without stirring. After ½ hour, the free water was discarded from the beaker. The change in the weight of the polymer due to dissolution or absorption of water was calculated.

Percentage weight change = 100 x (Initial weight of polymer – polymer after losing/gaining water)/Initial weight of polymer

25 ml of water with pH 5 was added and after ½ an hour, free water was discarded. The percentage change in weight was then calculated. These processes were continued until the polymer was completely dissolved. Similar measurements were carried out at different values of water concentration (pH 7.3 and pH 9.5). Percentage of weight changed due to absorption and dissolution of copolymer (FA-AA) in water at different times was measured and was plotted in Figure 1.

Figure1: Water absorption and dissolution rate of copolymer (FA-AA) in acid and alkali solution

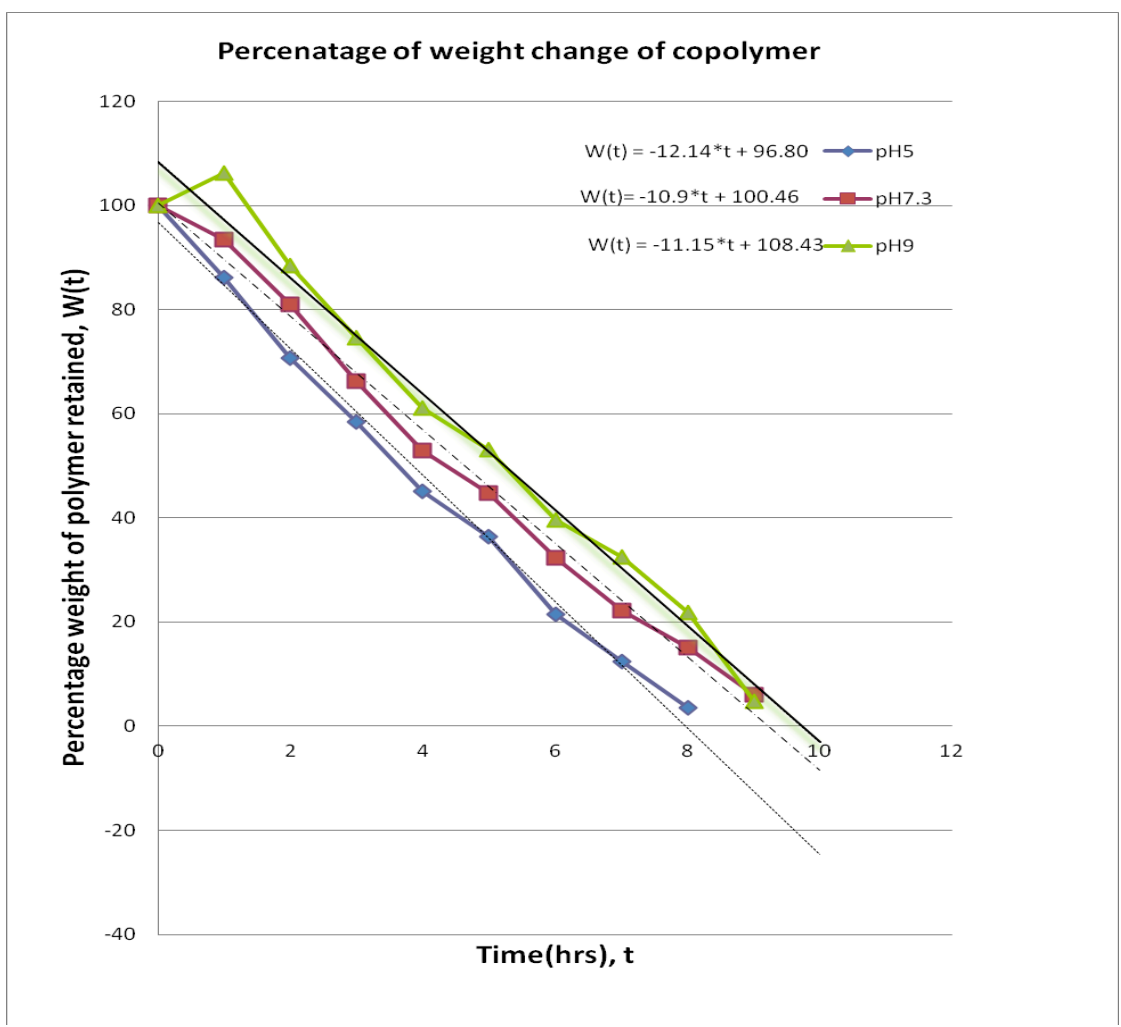

Figure 1 shows the percentage weight changes due to absorption and dissolution of copolymer (FA-AA) in water with changes in the pH of the water solution. The blue, red and green lines respectively indicate percentage weight decays with time for pH 5, 7.3 and 9.5. The figure clearly indicates an initial hike in absorption for pH 9.5 at an initial time scale of 2 hours, followed by an immediate decrease at all subsequent times, essentially leading to the same average decay rate for all pH concentrations (as indicated in the figure). As found in figure 1, the copolymer is completely dissolved in acid, tap water and alkali solution within 9 hours, 8 hours and 10 hours respectively. This shows that the solubility rate in acid solution (pH5) is higher than in any other solution and this may be due to the hydrolysis and a subsequent breakdown of the copolymer chain in acid solution. Remarkably, least square data fits from this figure (dotted lines for each of the three pH value plots) indicate approximately identical decay rates (11.15, 10.9 and 12.13 respectively for pH 5, 7.3 and 9.5) irrespective of the pH value. In a following section, an elementary rate kinetics based theory will be empirically proposed based on this result.

### 3. Results and Discussion

### 3.1 Water swelling and solubility characteristics of tercopolymers at different pH

Water absorption and solubility characteristics of tercopolymer (AA-AM-MMA) were tested using the same procedure as that of copolymers (FA-AA) at different pH. Data representing the percentage changes in weight are shown in Figure 2. The same figure indicates that tercopolymer is insoluble in acid solution but soluble in strong alkali solution. At pH 3, the copolymer absorbed water up to $1^{1}/_{2}$ times its weight for a period of 2 hours and thereafter it became saturated. The percentage weight gain remained nearly constant and the polymer remained in an insoluble form. Similarly at pH 5, the polymer absorbed more water than for pH 3 and then the percentage weight gain remains unchanged with the polymer remaining in an insoluble form. With a strong acid solution, carboxylic and ester group of tercopolymer formed a hydrogen bond by absorbing acid and become very inert with the weight change remaining practically constant. At pH 7, tercopolymer gradually absorbs water for a period of 12 hours. At pH 9.5, acid functional groups were neutralized and ester groups were hydrolyzed with water molecules penetrating inside the polymer matrix. Aided by this underlying phenomenology, the tercopolymer gained weight for a certain period and then dissolved slowly after 4-5 hours. The percentage of weight gain and loss were recorded to be faster at strong alkali solutions at and above pH 11.

Figure 2: Water absorption and dissolution rate of tercopolymer in acid and alkali

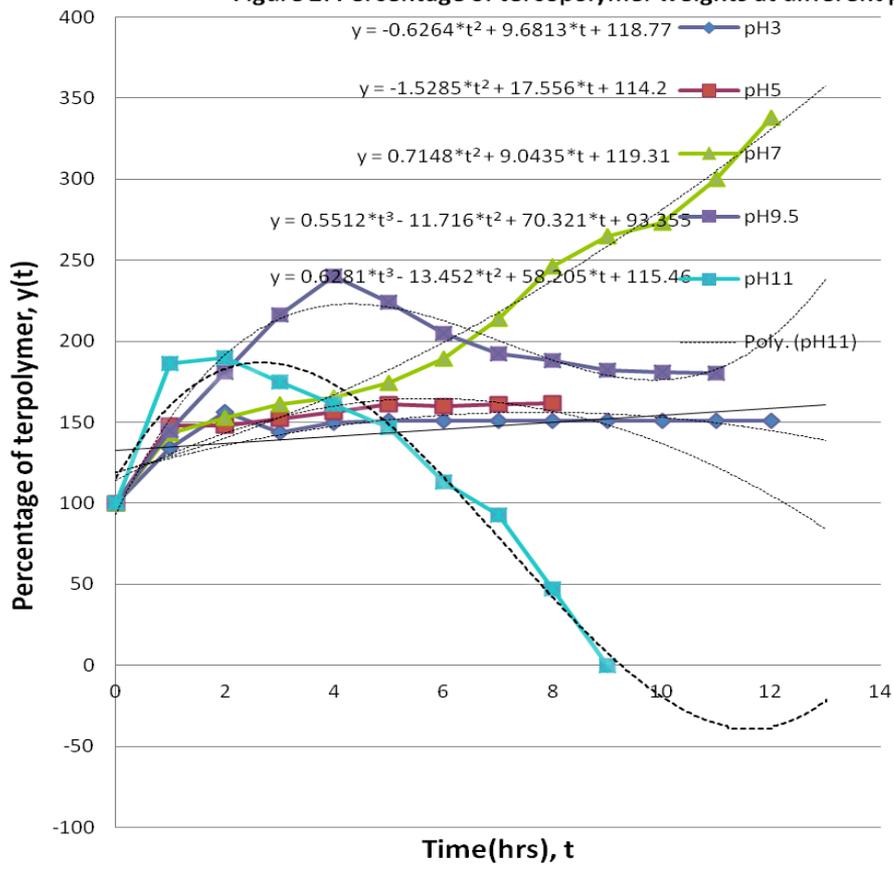

Figure 2: Percentage of tercopolymer weights at different pH

A noted difference between figures 1 and 2 is the nonlinearity dominance of tercopolymer decay rates. While the blue, red, green, purple and blue lines indicate pH values 3, 5, 7, 9.5 and 11 respectively, their respective least square (polynomial) fits abide the respective extrapolation relations:

- $y = -0.6264*t^2 + 9.6813*t + 118.77$ (pH 3)
- $y = -1.5285*t^2 + 17.556*t + 114.2$
- $y = 0.7148*t^2 + 9.0435*t + 119.31$ (pH 7)
- $y = 0.5512*t^3 - 11.716*t^2 + 70.321*t + 93.355$ (pH 9.5)
- $y = 0.6281*t^3 - 13.452*t^2 + 58.205*t + 115.46$ (pH 11)

**3.2 Solubility characteristics of tercopolymer by pH step change**

As the stomach in our body is acidic in nature and the intestine has an alkaline base, when polymer bound drugs enter into the stomach of our body, polymer swells in the body fluid and starts to release the drugs. In these experiments, we studied the water absorption and release characteristic from the tercopolymer by the step change method to match the release of drugs in similar conditions for a range of pH values, spanning the range of acidic to alkaline spectrum.

**3.2.1 Step change methods**

2 grams of tercopolymer (AA-AM-MMA) was taken in a beaker containing 40 ml of alkali solution at pH 8.5. The alkali solution was added and left to react with the polymer for 2 hours. After 2 hours, excess water lying above the surface of the tercopolymer was decanted from the beaker and washed with about 5 ml of distilled water. The weight of polymer was measured. 40 ml acidic solution at pH 5 was added again to the beaker containing tercopolymer and left for 2 more hours. The excess acidic water (pH 5) was decanted from the beaker and was washed with 5 ml distill water. The change in weight of the polymer was measured. 40 ml of alkali solution at pH 8.5 was added again and kept for 2 hours and subsequently the excess water with pH 8.5 was decanted and washed with 5 ml of distilled water. The change in weight of the polymer was measured. This process was continued once at pH 8.5 and subsequently at pH 5 for three times and the change in weight of the polymer was calibrated according to pH strength. Similar step change processes were carried out using 2 grams of tercopolymer on alkali solution at pH 9.5 and acidic solution at pH 5 and repeated thrice over.

Results of percentage weight gain or loss in step change experiments were plotted against time. As shown in Figure 3, tercopolymer absorbs water by neutralizing the acid group and by increasing the inter-molecular spacing. Tercopolymer absorbed more water at pH 9.5 than at pH 8.5. Higher alkaline water neutralizes more effectively and penetrates into the polymer matrix than at lower pH. When tercopolymer is immersed in an acid solution at pH 5, hydroxyl groups are neutralized by proton. The intermolecular space between the polymer molecules is reduced forming hydrogen bond together with release of water. Figure 3 also shows that tercopolymer absorbs more water in the step change process after 2-3 steps and releases more water. This is more prominent at pH 9.5 than for pH 8.5.

Figure 3: Solubility characteristics of tercopolymer in solution by pH step change

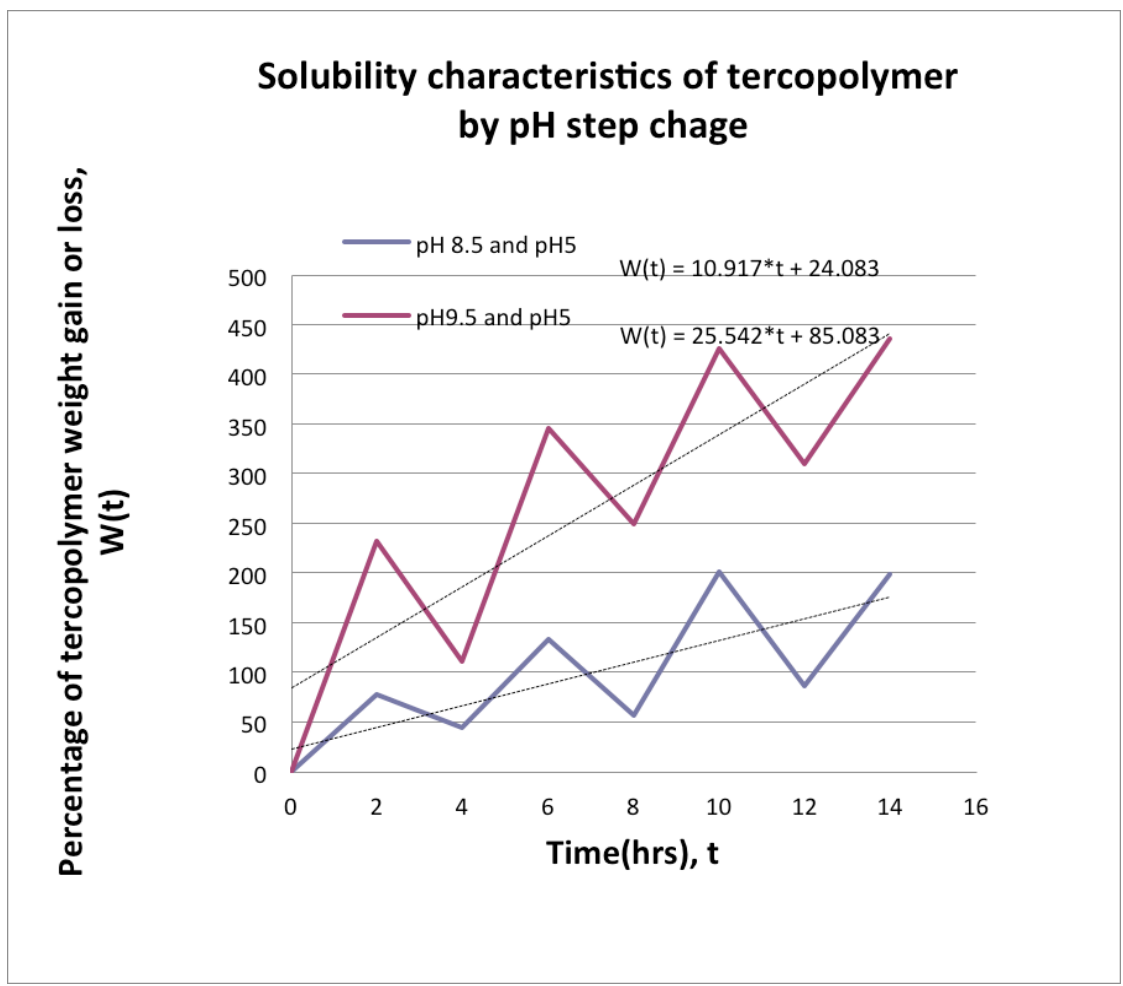

Figure 3 suggests a sinusoidal time dynamics of the solubility characteristics of percopolymer as a function of the pH of the solution. The mean envelopes (growth profile), though, are linearly correlated as shown in the figure above.

### 3.3 Biocompatibility

Microorganisms S.cocus and E.coli were inoculated in planting media such as agar and agar containing 20% polymers like FA-AA, AA-AM-MMA, PU in four different plates. The plates were incubated at $37^0C$ for 36 hours. Antibiotic drug cephalosporin was then applied into the plates and incubated at $37^0C$ for 24 hours. After this incubation period, all the plates were examined for inhibition of microorganisms` growth. The results of inhibition showed that the rate of inhibition were nearly same. The results also demonstrated the compatibility of media where the microbiology can grow and can be inhibited by antibiotics.

### 3.4 Controlled Release of antibiotics

In this study, 0.1 gm of antibiotic drug cephalosporin (Alembic pharmaceuticals Ltd, India) was blended with 2 grams of copolymer (FA-AA) and tercopolymer (AA-AM-MMA, PU-FA-AA) blend polymers respectively. Gram strained inoculated plates were developed in agar. The plates were incubated at $37\ ^0C$ for 24hours. 0.1 gm of polymer samples containing cephalosporin was then applied on the surface of the plates using needle dispenser. The plates were incubated at $37^0C$ for 24 hours. After the incubation, the diameter of the zone of inhibition was interpreted. The rates of inhibition of bacteria in different antibiotic bound polymers are shown in the Figure 4. The figure shows the rate of release of antibiotics drug from the polymer matrix and its rate of action to inhibit the growth of microorganism. The rate of release of drugs appears to be high in case of copolymer (FA-AA) (B. Bubling in Figure 4) as well as in tercopolymer (AA-AM-MMA) (S in the Figure 4) and low in PU-FA-AA and PU- tercopolymer blend ( B1 in figure 4) comparatively.

Figure 4: The rate of inhibition of bacteria from antibiotic bound polymers

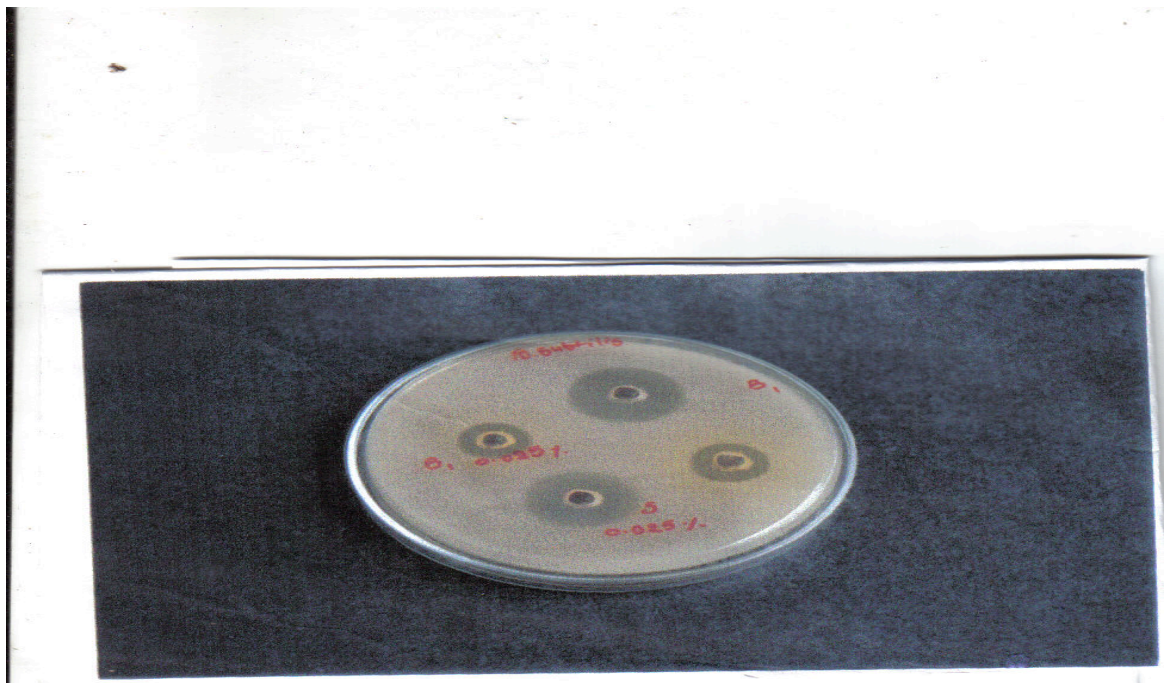

**3.5 Controlled release of paracetamol**
To study the suitability of using tercopolymers, PU-FA-AA, PU-Tercopolymer in controlled drug release, paracetamol 15% in weight was loaded in polymers solution and mixed well. The drug-loaded polymers were evaporated to dryness. Paracetamol was released at $37^0$C using dissolution tester (Dissotest, LabIndia, Mumbai, India). 1 gm of drug blend polymers having size 7-8 mm and 900 ml of buffer solution (pH8) were taken into the dissolution tester and stirred at 50 rpm. A 10ml of aliquot was taken each time for analyzing the paracetamol content at hourly intervals. The amount of paracetamol release was determined from the standard curve of paracetamol concentration (mg/L) versus absorbance. The absorbance was measured at 270nM in a spectrophotometer by adding reagents sodium hypochloride and alkaline sodium salicylate to paracetamol in serial dilution. The results of the paracetamol release from the different polymer matrices are shown in Figure 5. This figure shows the degree of release of paracetamol as a function of time in the order of TR-P> PU-FA-AA-P >PU-TR-P.

The release of paracetamol drugs from TR-P was much higher than PU-FA-AA-P and PU-TR-P at pH 8. 80% of the drugs were released within 5-6 hours. Mild alkaline water molecules interact directly with the polar group of tercopolymer and helped to release the drug much faster. 50% paracetamol was released from PU-FA-AA-P within 4 hours as compared to 8 hours for PU-TR-P at pH 3. This suggests that FA-AA copolymer leaks faster through the polyurethane matrix in acid solution and releases paracetamol. The drug release was found to be slow initially but sped up after 40-50% release of the drug. The cumulative release of drug was 90% in 8 hours for PU-FA-AA-P as compared to 12 hours for PU-TR-P at pH 3. The release of drugs was marginally higher for TR-P at pH3 when compared to PU-TR-P at pH8. The overall release of drugs was higher at pH 8 than at pH 3.

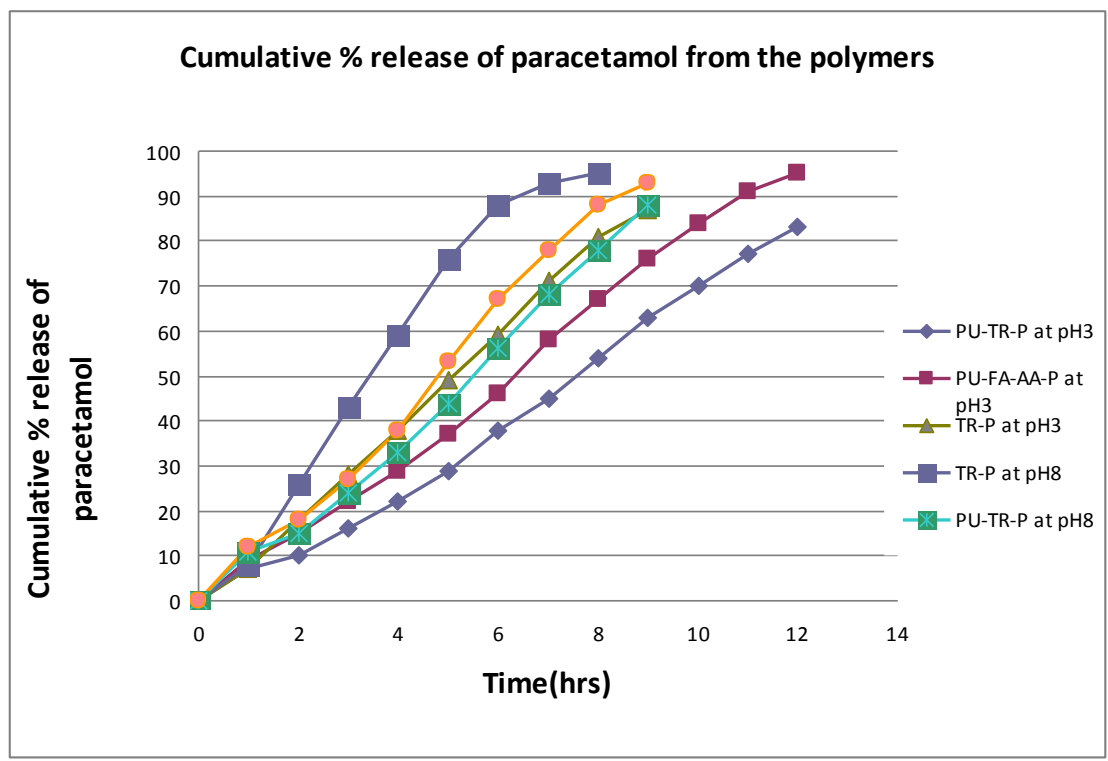

Figure 5. Rate of release of paracetamol from the polymers at different pH

Work is presently underway to reproduce and predict such rate kinetic statistics based on a Michaelis-Menten theory.

## 4 Conclusions

It is important that drug reaches the location of disease at a constant medically prorated concentration for maximum efficacy. The variation of therapeutic uncontrolled dosage might either be toxic or ineffective, the latter, in essence, being the equivalence of prolonged suffering. Controlled release using targeted drugs is the new route towards drug formulation based on advanced polymeric systems that are actively monitored based on continued recent research interests to deliver drugs to the site of the disease.

For these reasons, new types of pH sensitive hydrophilic nature polymers such as copolymer (fatty acid-acrylic acid) and tercopolymer (polyacryamide, acrylic acid and MMA) were developed and mixed within the matrix of polyurethane. The oil modified polymers developed from natural biological low cost resources showed a typical 9-10 hour decay time in an atmosphere similar to stomach condition pH 5-9, which could further be regulated by control polymerization. The tercopolymer showed a large decay time more than 12 hours at varied pH and showed dynamic swelling and weight loss characteristics at high and low pH (step change method). The polymer blend showed slow release of drugs favoring gram strained media as well as in pH 3 and pH 8. The release characteristics of drugs at different pH solutions were attributed to morphological and swelling characteristics of oil modified copolymer and tercopolymer and on its blends with polyurethane polymer that may be further regulated depending on the pH and polymer blend-ratio.

Remarkably, percentage weight loss as a function of time for copolymers recorded linear decaying profiles with the same mean decay rate irrespective of the pH of the solution whereas a strongly nonlinear profile was noted for the case of tercopolymers. Nonlinear polynomial-fitting (Figure 2) of the tercopolymer decay plots suggest the dominance of a nonlinear regime in the underlying phenomenology governing the decay rates. The results (Figure 3) can be compared with a sinusoidal increase of solubility with time for tercopolymers with a mean least square fitted envelope predicting a pH affected growth rate (as shown in Figure 3). The effects of paracetamol on the pH affected solubility profiles (Figure 5) define a stark contrast to that of the profiles in the absence of paracetamol (Figure 3) in that the former clearly indicates the presence of a 'saturation zone' where beyond a certain critical time, that is a function of the pH itself, the cumulative release of paracetamol remains unaffected with longer study periods. Efforts are presently underway to develop a probabilistic mathematical model that is capable of predicting future experimental results as functions of ambient parameters and the drug specifics within pre-specified error bounds.

## 5 Acknowledgements
This work is attributed to the Fuel cell, Department of Chemical Engineering, University of Birmingham group. NP is very grateful to Dr Marlin Thomas, ISTAR, India for her valuable support in arranging chemicals and for associated testing of samples.